# Relativistic Gross-Pitaevskii equation and the cosmological Bose Einstein Condensation

## -Quantum Structure in Universe-

Takeshi FUKUYAMA and Masahiro MORIKAWA†

*Department of Physics, Ritsumeikan University*
*Kusatsu Shiga, 525-8577, Japan*

†*Department of Physics, Ochanomizu University,*
*2-1-1 Otsuka, Bunkyo, Tokyo,112-8610, Japan*

Abstract

We do not know 96% of the total matter in the universe at present. In this paper, a cosmological model is proposed in which Dark Energy (DE) is identified as Bose-Einstein Condensation (BEC) of some boson field. Global cosmic acceleration caused by this BEC and multiple rapid collapses of BEC into black holes etc. (=Dark Matter (DM)) are examined based on the relativistic version of the Gross-Pitaevskii equation. We propose (a) a novel mechanism of inflation free from the slow-rolling condition, (b) a natural solution for the cosmic coincidence ('Why Now?') problem through the transition from DE into DM, (c) very early formation of highly non-linear objects such as black holes, which might trigger the first light as a form of quasars, and (d) log-z periodicity in the subsequent BEC collapsing time. All of these are based on the steady slow BEC process.

# 1. Introduction

It is amazing that recent cosmological observations provide us a wide rage of knowledge and mysteries. It is also amazing that the standard $\Lambda$CDM cosmological model works perfectly without specifying the most of the matter contents of the universe. In this model, the basic concepts and the structure of matter and space are both very simple, and the basic assumptions are clear. Especially within this model, the temperature fluctuations $\delta T$ in the sky and the large scale power spectrum $P(k)$ of density fluctuations can be perfectly calculable from the primordial density fluctuations [1];they correctly describe the most of the observations.

However as a theory of Physics, there are at least two unsatisfactory points in this standard model. (a) One is that the theory lacks the identification of matter. Although quite an amount of unknown matter plays an important role in the theory, they are simply called Dark Matter and Dark Energy and specifications of them are postponed. Thus we still don't know 96% of the cosmic contents: Dark Energy(DE) and Dark Matter(DM). Moreover we don't know any relation between them. (b) Another is that the successful description of and the harmony with the cosmological observations are only limited to the linear stage. There are many peculiarities in the Non-linear regime: too early formation of objects and re-ionization at around $z \approx 20$, the physical detail of the biasing for galaxy formation, and a natural mechanism how the first stars formed, etc. These facts force us to consider some other source of instability, as well as the ordinary gravity, in order to form clearly localized structures.

In this article, we would like to improve the standard $\Lambda$CDM model mainly based on the above two points. We have to bear the following basic facts in mind for that purpose. (A) The dominant DE should have negative pressure ($p < 0$) in order to make the cosmic expansion accelerate rather than decelerate. Actually from the Einstein equation for the scale factor $a(t)$ of the universe,

$$\ddot{a}(t) = -\frac{4\pi G}{3}(\rho + 3p)a(t), \tag{1}$$

the condition $\rho + 3p < 0$ must hold. This implies the dominance of very peculiar matter which cannot be described by the acquainted classical theories. (B) Various observations [1] indicate that the amount of DE and DM are almost the same, about 3:1. This fact indicates that they are intimately related with each other, possessing probably the common root.

Among familiar classical matters, such as gas, liquid, solid, plasma, etc., it is very difficult to find matter which shows negative pressure. The only matter which shows negative pressure would be the coherent scalar field $\phi$ [2]. In this article, we would like to specify the physical origin of this matter. Especially we emphasize that this scalar field may be the macroscopic wave function of the Bose-Einstein condensation (BEC) of some boson field with self interaction. According to the Bogoliubov prescription, the condensate $\phi$ of the field in the momentum space is clearly distinguished from the quantum part $\delta\hat{\phi}$:

$$\hat{\phi} = \phi + \delta\hat{\phi} \tag{2}$$

We propose a model in which Bose Einstein Condensation (BEC) of some bose field works as DE[3]. Moreover, DM will be naturally realized as locally collapsed BEC.

In our previous paper [3], we actually proposed the cosmological model of this type based on the Bose-Einstein condensation. There the condensate was described by the ordinary non-relativistic Gross-Pitaevskii (GP) equation:

$$i\hbar\frac{\partial \psi}{\partial t} = -\frac{\hbar^2}{2m}\Delta\psi + V\psi + g|\psi|^2\psi. \tag{3}$$

Here $\psi(\vec{x}, t)$ is the condensate mean field, and $V(\vec{x})$ the potential. Also $g = 4\pi\hbar^2 a/m$, and $a$ is the s-wave scattering length. However, this non-relativistic GP equation is not sufficient to describe the extreme regime of the state equations such as $p = -\rho$ which we assumed. Especially the zero-point of the energy is ambiguous in the non-relativistic treatment. This point is essential to keep consistency with Einstein equation in which the full energy contributes[1]. For a reliable argument, we need the relativistic version of GP equation and must argue BEC based on it without phenomenological equation of state such as $p=-\rho$. Thus in this paper, we will extend our previous scenario based on the relativistic version of GP equation and properly consider the BEC dynamics based on it.

After examining the possibility of and the basic conditions for BEC in the universe in section 2, we derive the relativistic GP equation in section3. Based on this equation and the Einstein equation, we will find two relevant regimes in the cosmic evolution in section 4. In section5, we examine the instability and the collapse of BEC suggested in the previous section. We then argue the possible predictions and the observational tests of our model in section 6. Section 7 is devoted to the summary and the relation of our results with other works.

## 2. Cosmic BEC mechanism

Let us first examine the problem under which condition BEC is possible in the expanding universe. In general BEC is possible if the thermal de Broglie length exceeds the mean separation of particles of the bose gas, $\lambda_{dB} \equiv \left(2\pi\hbar^2/(mkT)\right)^{1/2} > r \equiv n^{-1/3}$. This condition determines the critical temperature below which BEC can initiate:

$$T_{\rm cr} = \frac{2\pi\hbar^2 n^{2/3}}{km}. \qquad (4)$$

On the other hand, the cosmic temperature of non-relativistic matter has the same dependence on the density!

$$T = T_0 \frac{2\pi\hbar^2}{m}\left(\frac{n}{n_0}\right)^{2/3}. \qquad (5)$$

Here we used the temperature dependence of the non-relativistic matter $\rho \propto a^{-3} \propto T^{3/2}$, which is different from that of radiation $\rho \propto a^{-4} \propto T^4$. These equations come from the entropy conservation during the expansion. Further, the boson must be non-relativistic for BEC since otherwise (if relativistic $kT > mc^2$) the particle number is not conserved and the ordinary BEC in the momentum space cannot be realized. From this fact and the same dependence on the density in Eqs.(4) and (5), we conclude that once the temperature becomes lower than the critical temperature and the mass,

$$T < T_{\rm cr} \quad \text{and} \quad kT < mc^2 \qquad (6)$$

at some moment of the cosmic expansion, then BEC has a chance to initiate all the time after that moment. A rough estimate of the condition is possible; if the boson temperature coincides with the radiation temperature at $z=1000$, then using the present matter density $\rho = \rho_{now} = 9.44 \ 10^{-30} {\rm g/cm}^3$, we have the BEC condition $m < 1.96 {\rm eV}$. However, this mass constraint is not decisive since the boson temperature needs not to be the same as the radiation temperature; for example the axion case [4].

There is yet another condition for BEC actually to take place in the universe. BEC is often claimed to be a phase transition without interaction. However if all degrees of freedom are exactly free from any interaction, then there would be no level transition and especially no condensation toward the lowest energy. Therefore, there must be any small interaction which makes the energy-level transition possible.

## 3. Relativistic Gross-Pitaevski Equation

As explained in the introduction, the relativistic version of the Gross-Pitaevskii equation is indispensable for the proper description of BEC consistent with the Einstein equation. The non-relativistic GP Eq.(3) is the form of the non-linear Schrödinger equation. The ordinary Schrödinger equation is the non-relativistic approximation of the Klein-Gordon equation. Therefore relativistic GP equation must be the Klein-Gordon equation with appropriate interaction.

---

[1] Furthermore, within the non-relativistic argument, the properties (a) and (d) in the abstract could not be derived.

Thus we have the relativistic version of GP equation

$$\frac{\partial^2 \phi}{\partial t^2} - \Delta \phi + m^2 \phi + \lambda(\phi^*\phi)\phi = 0, \tag{7}$$

with the potential

$$V \equiv m^2 \phi^* \phi + \frac{\lambda}{2}(\phi^*\phi)^2. \tag{8}$$

This Eq.(7) is derived from the Lagrangian

$$L = \sqrt{-g}\left[g^{\mu\nu}\partial_\mu\phi^* \cdot \partial_\nu\phi - m^2\phi^*\phi - \frac{\lambda}{2}(\phi^*\phi)^2 + L_g\right]. \tag{9}$$

Substituting the decomposition $\phi = Ae^{iS}$ and defining $p_\mu = -\partial_\mu S = (\varepsilon, -\vec{p})$, where $\vec{p} = m\gamma\vec{v}$, $\gamma = (1-(\vec{v}^2/c^2))^{-1/2}$, the relativistic GP equation reduces to the Euler equation for fluid:

$$\varepsilon \frac{\partial \vec{v}}{\partial t} + \vec{\nabla}\left(\frac{\gamma v^2}{2} + \frac{\lambda}{12m}A^2 + \frac{\hbar^2}{2Am}\Box A\right) = 0. \tag{10}$$

The energy-momentum tensor of Eq.(9) becomes

$$T_{\mu\nu} \equiv \frac{2}{\sqrt{-g}}\frac{\delta L}{\delta g^{\mu\nu}} = 2\partial_\mu\phi^*\partial_\nu\phi - g_{\mu\nu}(\partial\phi^*\partial\phi - m^2\phi^*\phi - \frac{1}{2}(\phi^*\phi)^2). \tag{11}$$

For the isometric relativistic fluid, it takes the form

$$T^{\mu\nu} = diag(\rho, p, p, p) \tag{12}$$

in the local rest frame. Here the condensate part of $\rho$ and $p$ are given by

$$\rho = T^{00} = \dot{\phi}^*\dot{\phi} + m^2\phi^*\phi + \frac{\lambda}{2}(\phi^*\phi)^2 = \dot{\phi}^*\dot{\phi} + V \tag{13}$$

and

$$p = T^{11} = T^{22} = T^{33} = \dot{\phi}^*\dot{\phi} - m^2\phi^*\phi - \frac{\lambda}{2}(\phi^*\phi)^2 = \dot{\phi}^*\dot{\phi} - V. \tag{14}$$

Now let us consider a spatially uniform solution to Eq.(7). Substituting $\phi = A_0 e^{i\omega t}$ into Eq.(7), we obtain

$$\omega^2 = m^2 + \lambda A_0^2 \tag{15}$$

and

$$\rho = (\omega^2 + m^2)A_0^2 + \frac{\lambda}{2}A_0^4, \tag{16}$$

$$p = (\omega^2 - m^2)A_0^2 - \frac{\lambda}{2}A_0^4. \tag{17}$$

Manipulating Eqs.(15)-(17), we obtain

$$\rho = 2m^2 A_0^2 + \frac{3\lambda}{2}A_0^4, \tag{18}$$

$$p = \frac{\lambda}{2}A_0^4. \tag{19}$$

It should be remarked that the pressure becomes negative $p < 0$ for $\lambda < 0$, which we assume, and

$$\rho + 3p = 2m^2 A_0^2 + 3\lambda A_0^4. \tag{20}$$

Namely, since $\lambda < 0$, the condensate of $A_0^2$ grows from 0 and when

$$\xi \equiv \frac{|\lambda|A_0^2}{m^2} > \frac{2}{3}, \tag{21}$$

it breaks the strong energy condition; the universe starts the accelerated expansion. Furthermore, the weak energy condition breaks down for $\xi > 1$ and the positivity of $\rho$ breaks down for $\xi > 4/3$.

Note that the mean fields $\psi$ and $\phi$ describing the BEC above are classical fields. Moreover the bose field we consider is most likely a composite field of fermion pair but not a fundamental field. Thus the negative signature of the coupling constant $\lambda < 0$ may be acceptable, contrary to the fundamental quantum case in which the vacuum becomes unstable for $\lambda < 0$.

# 4. Steady slow process of BE-condensation

As explained in section 2, the speed of the BEC is controlled by the strength of the interaction which makes the energy-level transition possible, especially the transition toward the ground state. Keeping this fact in mind, we suppose that the BE-condensation slowly proceeds with steady speed $\Gamma$ [2]. Then rewriting the Einstein equation for the FRW universe and the relativistic GP equation on this metric, our basic set of equations which describe the temporal evolution of densities becomes

$$H^2 = \left(\frac{\dot{a}}{a}\right)^2 = \frac{8\pi G}{3c^2}(\rho_g + \rho_\phi),$$
$$\dot{\rho}_g = -3H\rho_g - \Gamma\rho_g, \quad (22)$$
$$\dot{\rho}_\phi = -6H(\rho_\phi - V) + \Gamma\rho_g.$$

Here $\rho_g, \rho_\phi$ are the bose gas and BEC energy densities, respectively. We would like to emphasize, in our model, that *the essence of the cosmic BEC model is the slow and steady condensation flow*.

There are two relevant regimes of solutions for Eq.(22). One is the over-hill regime (a) and another is the inflationary regime (b). The former appears when the condensation speed is too high and the latter too low. Let us now examine each of them.

(a) *Over-hill regime*: This regime appears when the condensation speed is faster than the potential force. This condition is generally thought to be realized in the earlier stage of the cosmic evolution. Actually this regime is a fixed point of the set of equations (22):

$$\phi \to \infty, \rho_\phi \to 0, \rho_g \to 0, H \to 0, a \to a_* \quad (23)$$

When the condensation speed $\Gamma$ is fast, the bose gas density simply reduces by this effect: $\rho_g \propto e^{-\Gamma t}$. The field goes over the hill of the potential, as in Figure 3, and it simply falls $\phi \to \infty$ finally. Then the friction term associated with the cosmic expansion also becomes negligible. Therefore the equation in Eq.(22) yields $\ddot{\phi} \approx V'$, from which we know that $\phi$ reaches singularity within a finite time interval. From the first line of Eq.(22), we have $H \propto \sqrt{\rho_g + \rho_\phi} > \sqrt{\rho_g} \propto e^{-\Gamma t/2}$. This property and the fact that $\dot\phi$ rapidly increases in the last stage of the fall, we conclude that the BEC reduction rate $-6H(\rho_\phi - V) \propto H\dot\phi^2$ dominates BEC increase rate $\Gamma\rho_g$. Thus eventually $\rho_\phi \to 0$, and $H \to 0$.

As is clear from Eq.(23), the cosmic expansion stops in this regime and a naive linear stability analysis is applicable. According to [5], the condensate $\phi$ behaves as

$$\ddot{\phi} - \Delta\phi + m^2(1 + 2\Psi)\phi + \lambda\phi^3 = 0, \quad (24)$$

where $\Psi$ is the gravitational potential. The linear analysis reveals that the Jeans instability sets in for $k < k_J$ such that

$$k_J^2 = -\lambda A_0^2 + \sqrt{(\lambda A_0^2)^2 + 16\pi Gm^2 A_0^2 \omega^2}, \quad (25)$$

where $A_0$ is the background value of $\phi$ and $\omega^2 = m^2 + \lambda A_0^2$ is defined in Eq.(15). Jeans wave length corresponding to this wave number is mostly determined by the mass $m$, which is supposed to be about eV. Since this is a microscopic scale, all the relevant fluctuation mode would become unstable.

Moreover, the weak energy condition is apparently violated, $-p_\phi > \rho_\phi$, after some point during the approach toward this regime. The violation of the weak energy condition deduces the violation of causality and the appearance of this violation indicates the inconsistency of the model, especially the assumption of the uniform distribution of the condensate. Thus the uniform mode of BEC should eventually collapse to form local

---

[2] This condensation rate should be derived from the fundamental interactions with many-body effects in the cosmic evolution environment. However the derivation would become lengthy, and therefore in the present model, we simply assume that this rate is some constant, as in the usual treatment.

structures.

Here in this paper we simply assume that the uniform BEC rapidly collapses into local structure, whose energy density $\rho_l$, with the speed $\Gamma'(t)$. Note that $\Gamma'(t)$ is non-zero only when the BEC becomes unstable. This phenomenological argument is sufficient for our present study, though the detailed derivation of the collapsing process would be definitely necessary for the precise prediction such as the mass function of the collapsed objects. Then our previous set of evolution equations Eq.(22) is modified as

$$H^2 = \left(\frac{\dot{a}}{a}\right)^2 = \frac{8\pi G}{3c^2}(\rho_g + \rho_\phi + \rho_l),$$
$$\dot{\rho}_g = -3H\rho_g - \Gamma\rho_g,$$
$$\dot{\rho}_\phi = -6H(\rho_\phi - V) + \Gamma\rho_g - \Gamma'\rho_\phi,$$
$$\dot{\rho}_l = -3H\rho_l + \Gamma'\rho_\phi,$$
(26)

where essentially the last line is newly added to Eq.(22). We have demonstrated a typical solution in the following three figures.

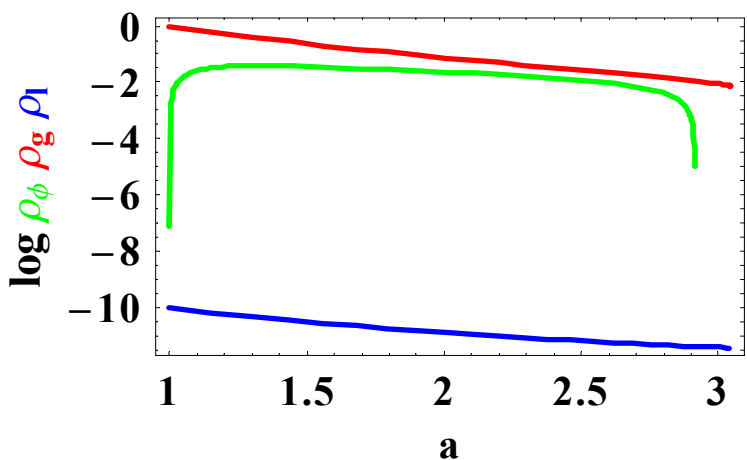

**Figure 1. Evolution of the energy densities in the over-hill regime. This is a numerical simulation of Eq. (26) for large condensation speed. Red and green curves are the bose gas and BEC energy densities, respectively ($\rho_g, \rho_\phi$) as a function of scale factor $a$, whose unit is arbitrary. Blue curve is the possible other energy density decayed from BEC. This and bose gas energy density behave as cold dark matter (CDM). The parameters for this numerical calculation is not special; we simply set** $m^2 = 0.01, \lambda = 0.1, \Gamma = 0.4, \rho_l^{\text{initial}} = 10^{-10}, \rho_g^{\text{initial}} = 1, \phi^{\text{initial}} = \dot{\phi}^{\text{initial}} = 10^{-4}$ **in the unit of** $8\pi G/(3c^2) = 1$. **We assumed that the BEC collapse is instantaneous;** $\Gamma' = \theta(-(\rho+p))c$, **where** $c$ **is a huge positive constant. In other words,** $\Gamma' = \infty$ **when the BEC violates the weak energy condition, and** $\Gamma' = 0$ **otherwise.**

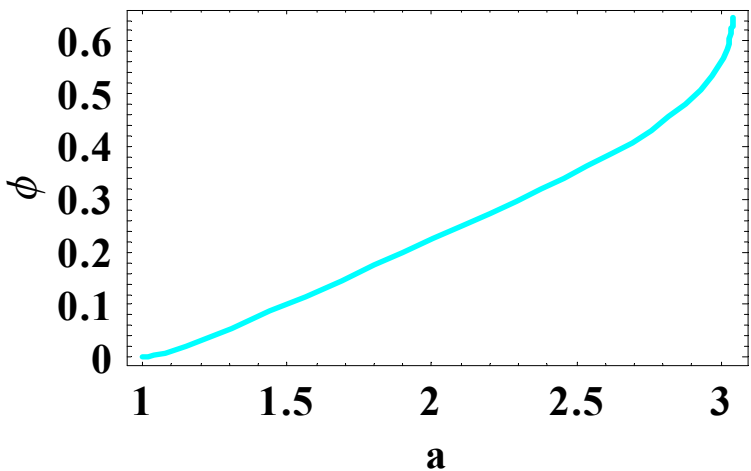

**Figure 2. Evolution of the BEC in the over-hill regime as a function of scale factor** $a$**, whose unit is arbitrary. It eventually goes beyond the potential hill (maximum of the potential** $V(\phi)$**) and falls into a**

singularity within a finite time interval. However actually before that the energy condition, and therefore the uniformity of BEC, breaks down. This inevitably leads to the BEC collapse.

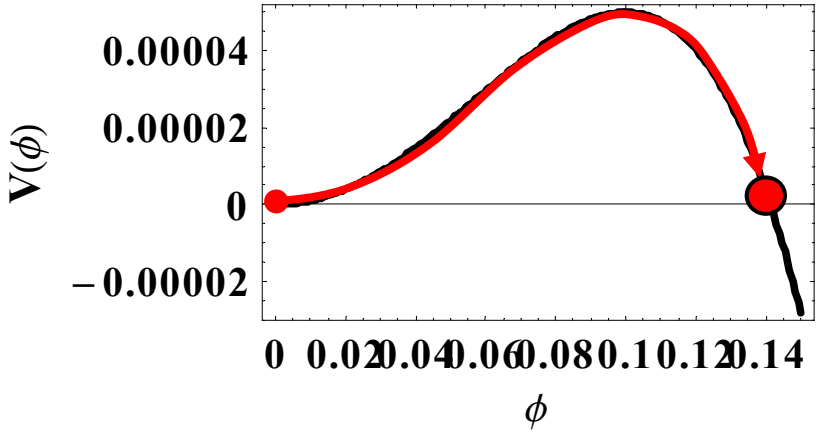

Figure 3. Schematic diagram of the evolution of BEC in the over-hill regime. The mean field $\phi$ goes beyond the potential hill.

(b) *Inflationary regime*: This regime appears when the condensation speed is slower than the potential force. This condition is generally realized in the later stage of the cosmic evolution. Actually this regime is a fixed point of the set of equations Eq.(22) or Eq.(26):

$$\phi \to \phi_*, H \to H_*, \rho \to 0, \dot{\phi} \to 0 \ . \tag{27}$$

Moreover, the linear analysis around this fixed point reveals that this is a stable fixed point; it has one zero-mode and one decaying mode.

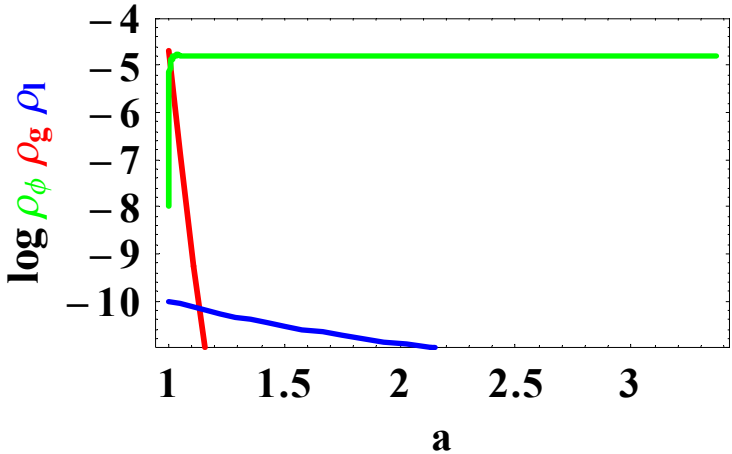

Figure 4. Same as Figure 1 but in the inflationary regime. This is a numerical simulation of Eq. (26) for small condensation speed. Parameters are the same as Figure 1 but $\rho_g^{\text{initial}} = 2 \times 10^{-5}$.

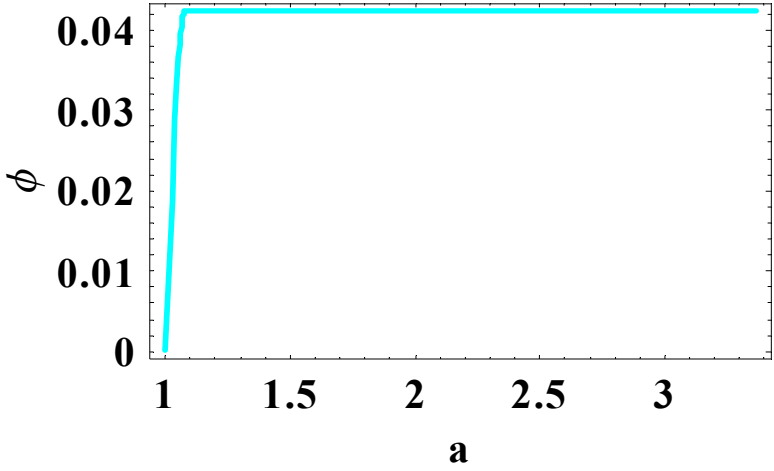

**Figure 5.** Same as Figure 2 but in the inflationary regime. The universe finally enters into this inflationary regime. The condensation speed becomes weak and the mean field $\phi$ cannot go beyond the potential hill but balances with the potential force.

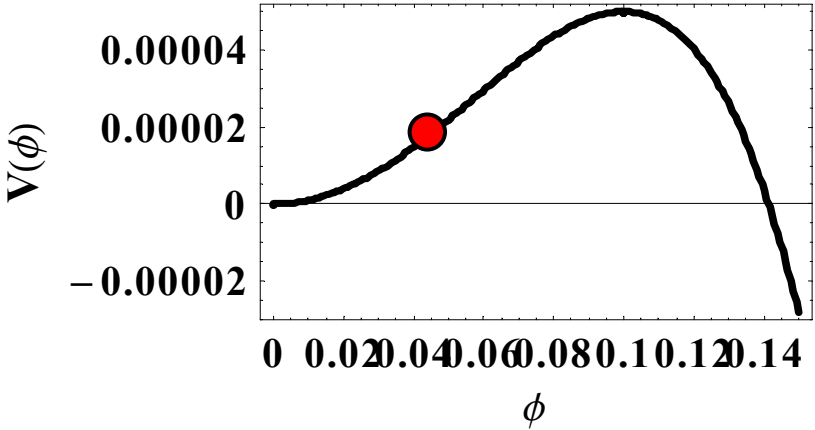

**Figure 6.** Same as Figure 3 but in the inflationary regime. The mean field $\phi$ stops at the middle of the potential hill. This vacuum energy induces the inflation. See Figure 7 in detail.

In this occasion, we would like to emphasize that this regime is a novel type of inflation. This inflation is supported by a balance of the condensation speed ($\Gamma \rho_g$), which promotes the condensation $\phi$, and the potential force ($\dot\phi V'$), which demotes the condensation $\phi$. Actually at the fixed point, the exact balance

$$\dot V = \Gamma \rho_g \tag{28}$$

is established. This can be checked by dividing the both sides of the last equation in Eq.(22) by $\dot\phi$ and applying the fixed point condition Eq.(27). Though the both sides of Eq.(28) exponentially reduce to zero, the balance itself is kept automatically[3].

Note that this novel inflation always takes place at the later stage of cosmic expansion irrespective of the detail of the initial conditions and the potential form. In this sense, this inflation is robust. On the other hand, the ordinary inflationary scenario requires the conditions $V'(\phi) \approx 0, V(\phi) \neq 0$, or more precisely, the slow-roll conditions

$$\frac{m_{pl}^2}{16\pi}\left(\frac{V'}{V}\right)^2 \ll 1, \quad \frac{m_{pl}^2}{8\pi}\frac{V''}{V} \ll 1, \tag{29}$$

---

[3] This exponentially reducing amplitude of the balance may lead to the instability of the inflationary regime and the autonomous termination of this regime, provided any small external perturbations.

from which the present novel inflation mechanism is free.

We can rephrase the above argument as *the steady and slow condensation supports the novel inflation which is robust and ubiquitous*.

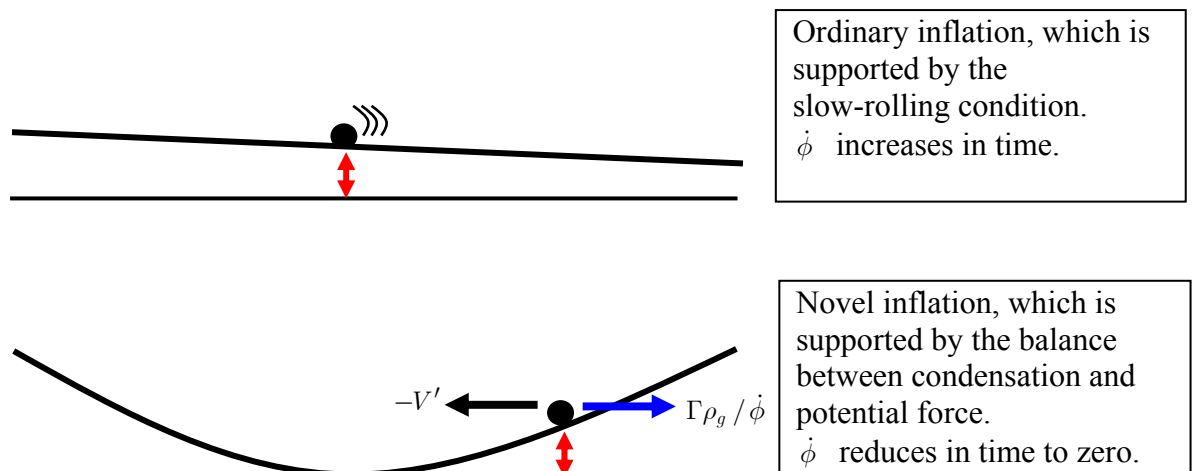

**Figure 7. Schematic comparison of the ordinary (above) and novel (below) inflationary mechanisms. The ordinary inflation is realized when the potential is very flat. The novel inflation is dynamically realized by the slow steady condensation for almost any potential. The necessary balance $\dot{V} = \Gamma \rho_g$, or in terms of forces $V' = \Gamma \rho_g / \dot{\phi}$, is automatically established. In this sense the novel inflation is quite robust and ubiquitous. The thick black lines represent typical potentials and the red lines represent the amount of vacuum energy in both cases.**

(c) *Several over-hill regimes finally followed by the inflationary regime.* (Figure 4-Figure 6): In the actual universe, the above two kinds of regimes realize subsequently. First, the over-hill regime repeats multiple times until the bose gas density reduces and the condensation speed reduces to balance the potential force, in which regime the final inflationary regime follows. In Figure 8, the evolution of cosmic energy densities is numerically demonstrated. Red, green, and blue curves represent the cosmic energy density of the bose gas ($\rho_g$), BEC ($\rho_\phi$), and the localized objects ($\rho_l$). In this example, four over-hill regimes are finally followed by the inflationary regime.

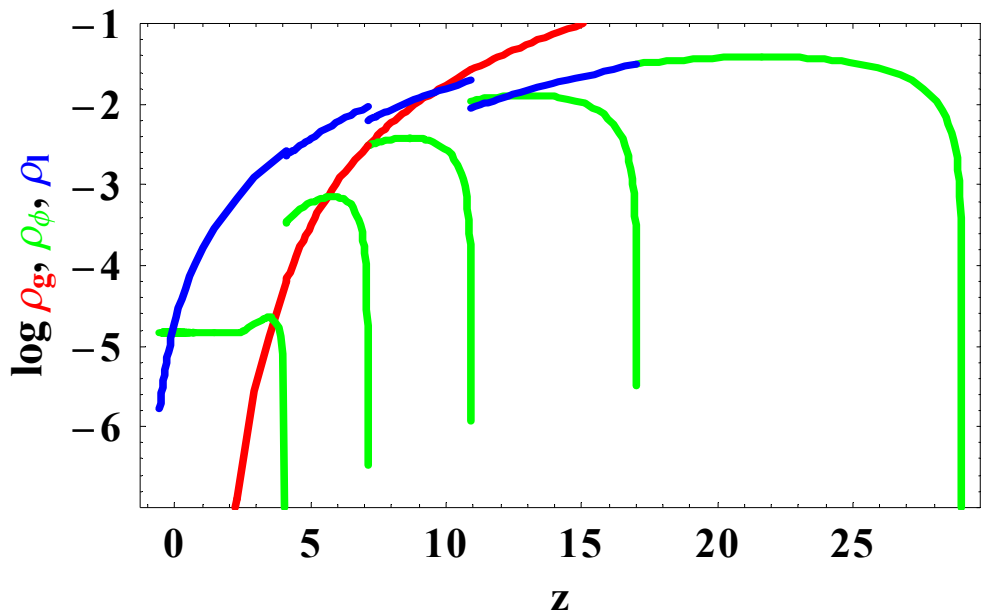

**Figure 8. A time evolution of various cosmic densities as a function of redshift. This is a numerical solution of Eq.(26); the combination of** Figure 1 **and** Figure 4. **Red, green, and blue curves are the cosmic**

energy density of the bose gas ($\rho_g$), BEC ($\rho_\phi$), and the localized objects ($\rho_l$), respectively. Four over-hill regimes are finally followed by the inflationary regime. Parameters are the same as Figure 1, and $\Gamma'$ instantaneously works immediately at the weak energy condition is violated. The densities of DE and DM are roughly the same order, however still requires some amount of fine tuning, which is not yet rendered in this demonstration, for the exact realization of the present ratio of DE/DM.

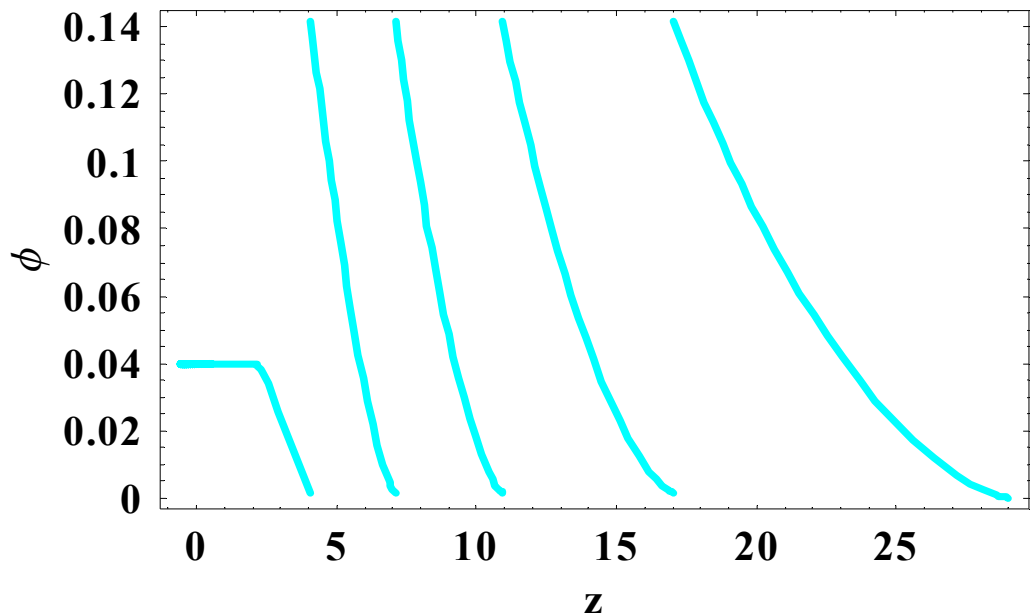

**Figure 9.** A time evolution of the condensate $\phi$ as a function of the scale factor. This is a numerical solution of Eq.(26); the combination of Figure 2 and Figure 5. The condensate goes beyond the hill four times and finally settles down to a constant value due to the automatic balance of the condensation flow and the potential force (novel inflation mechanism).

As is easily seen from Eq.(26),, DE($\rho_c$) and DM($\rho_g + \rho_l$) are intimately related with each other in our model; the amount of DE collapses into localized objects ($\rho_l$) which eventually becomes the dominant component of the total DM ($\rho_g + \rho_l$). Therefore, it is natural to expect that the amounts of DM and DE are almost the same, especially in the later stage of the cosmic evolution. This can be a kind of self organized criticality[6] (SOC) realized in the universe in the sense that one phase (DE) subsequently becomes critical state and rapidly transforms into a new phase (DM). This is true in the regime of repeated over-hill modes. However, once the inflationary regime begins, almost constant DE eventually dominates the diluting DM, and SOC disappears hereafter. Although this DE dominance in the late stage of the cosmic expansion is actually realized in our universe, some amount of fine tuning is still required for the exact realization of the present ratio of DE/DM.

In the context of BEC collapse, we would like to point out log(t)-periodicity in the collapsing times. As already pointed out, in the over-hill regime, the condensation speed $\Gamma$ is fast and the bose gas density simply reduces by this effect: $\rho_g \propto e^{-\Gamma t}$. Well before the field $\phi$ falls into singularity and $\dot\phi$ is small compared to the potential, the BEC reduction rate $-6H(\rho_\phi - V) \propto H\dot\phi^2$ is negligible compared to the BEC increase rate $\Gamma \rho_g$ [4]. This means $\rho_g$ is simply transformed into $\rho_\phi$ almost without cosmic dilution. We expect each BEC collapse takes place when $\rho_\phi$ ever reaches to the critical value $\rho_\phi^{\text{cr}} = V_{\max} = m^4/(-2\lambda)$ which is determined by the maximum height of the potential hill. Just after each collapse, new BE-condensation initiates from $\phi = 0$. Then the condensation energy density approximately behaves as

$$\rho_\phi(t) \approx [\rho_g(t_0) - \rho_g(t)]_{\text{mod } \rho_\phi^{\text{cr}}} \approx [\rho_g(t_0)(1 - e^{-\Gamma t})]_{\text{mod } \rho_\phi^{\text{cr}}} \tag{30}$$

---

[4] This situation is opposite to the violent case just around the over-hill fixed point.

where $t_0$ is the time when the condensation begins[5]. Accordingly we expect that each BEC collapse takes place after the time interval $\Delta t$ from the preceding collapse at time $t$, where $\Delta t$ is determined by the condition.

$$\rho_\phi^{\rm cr} = \rho_g(t_0)\left(e^{-\Gamma t} - e^{-\Gamma(t+\Delta t)}\right) \tag{31}$$

This implies that the occurrence of the BEC collapse is periodic in the logarithm of time $\log(t)$. Especially for not very late stage $z > 1$, and the cosmic expansion is almost power law in time $a(t) \propto t^{\rm const.}$, log(t)-periodicity means log(a) and log(z) -periodicities. For example in the numerical calculation in Figure 8, BEC collapse takes place at $z = 17.0, 10.9, 7.1, 4.1$, which is almost log-periodic.

Present log-periodicity in our model is similar to the ordinary one actively studied in the field of earth science and complex systems [7]. However one big difference is that we have no asymptotic accumulation point while the ordinary one often accompanies such point, which corresponds to the catastrophic singular point[6] [8]. Another difference would be the origin of the periodicity. In our case, the exponentially changing (either decaying or growing in general) flow ($\Gamma \rho_g$) and the finite constant capacity ($\rho_\phi^{\rm cr}$) of it leads to the log-periodicity. On the other hand in the ordinary one, the complex characteristic index of certain phase transitions is responsible for the log-periodicity.

This log(z)-periodicity is mostly general consequence of our model and these periodic BEC collapse may leave its trace in non-linear regime such as the discrete scale invariance or the hierarchical structure in the universe. The detailed argument will be the future problem.

---

[5] Incidentally, one may think that the expansion rate at the final inflation $\bar{H}$ is approximately determined by the value of Eq.(30) evaluated at $t \to \infty$: $\bar{H}^2 = \frac{8\pi G}{3c^2}[\rho_g(t_0)]_{{\rm mod}\,\rho_\phi^{\rm cr}}$. However, this is too rough since the values of $\rho_\phi^{\rm cr}$ and $\rho_g(t_0)$ are different with each other in many decades, and the cosmic dilution effect dominates in the early stage.

[6] Therefore, for example in the earth science, the prediction of the earthquake time is possible by finding the log-periodicity and its accumulation point in the change of the ground level or crust strain.

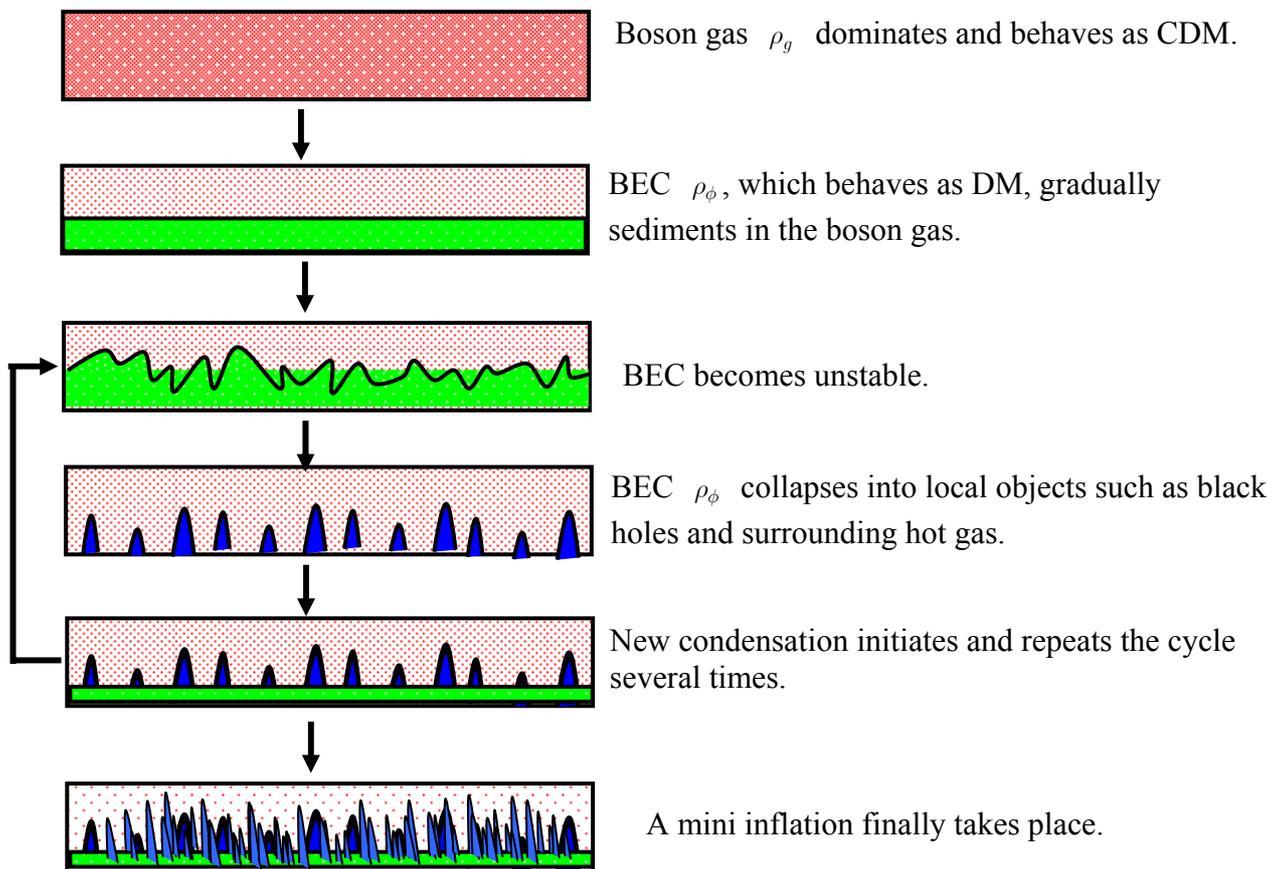

**Figure 10.** Schematic evolution of the cosmic energy densities in the BEC cosmological model. *Boson gas $\rho_g$ gradually condensates into $\rho_\phi$, which cause intermittent violent collapse into localized objects $\rho_l$.*

## 5. Collapse of BEC

Let us now examine the collapsing process of BEC in detail. We first study the first linear stage (a) and then the final non-linear further collapsing stage (b).

*(a) The linear stage*
After the dominance of $\rho_\phi$, the linear perturbation equation for the gauge invariant quantity yields [4]

$$\ddot{\delta}_k + 2H\dot{\delta}_k = \left[3H^2 + 2\left(\frac{k}{a}\right)^2\right]\delta_k , \qquad (32)$$

where we have used $c_s^2 \equiv \partial p/\partial \rho = \lambda\rho/(2m^2)$, $w \equiv p/\rho = \lambda\rho/(4m^2)$ and $\lambda < 0$, in the extreme region $-\lambda\rho \approx 4m^2$ (i.e. $p \approx -\rho$) just before the fixed point Eq.(23). Since $c_s^2 < 0$, there is no sound wave. Then, small scale mode ($k \gg aH$) becomes extremely unstable,

$$\delta_k \propto a(t)^{k/aH} \qquad (33)$$

and grows very rapidly. Since this collapsing time can be very fast compared to the condensation time scale $1/\Gamma$, the above approximation, $c_s^2, w$ are almost constant, is justified. At first glance, above rapid BEC collapse could violently destroy most of the results of the standard $\Lambda$CDM model. However this is not true as we will explain in the next section.

*(b) Further collapse*
For bosons, there is no degenerate pressure like fermions. Only the Heisenberg uncertainty principle or the quantum pressure, expressed in the last term in the left hand side of Eq.(10), can work to prevent the BEC to collapse. In order for this to happen, the Compton wave length (which should be the smallest size of the object) $\lambda_{comton} = 2\pi\hbar/(mc)$ should be larger than the Schwarzschild Black Hole radius or its inner-most stable radius

$3(2GM/c^2)$. This condition yields the critical mass

$$M_{critical} \approx m_{pl}^2/m \equiv M_{KAUP} \qquad (34)$$

only below which a stable configuration is possible. This structure is the boson star[9]. For example, the boson mass $m = 10^{-5}$ eV gives $M_{critical} = 10^{-5} M_\odot = M_\oplus$, which is almost the planetary mass. If the object mass exceeds this critical value, black holes are inevitably produced. Therefore, many compact clumps (boson stars and black holes) are rapidly formed after the collapse of BEC at the smallest scale.

However the BEC collapse is not so simple. During the collapse of BEC, the gravitational potential energy $GM^2/R$ can be released. If the collapse is free from the shock wave, then the most of the collapsing BEC would turn into black holes. On the other hand if the shock wave is produced, then this amount of energy is transformed into thermal energy to heat up the condensation. Roughly estimating,

$$NT \approx GM^2/R \approx Nm \qquad (35)$$

using the virial theorem, the temperature of the condensate, can rise to the mass of the boson at around the Schwarzschild radius $R \approx 2GM$. This means that the boson becomes relativistic and the particle number is no longer conserved. Then the basic condition for the BEC is lost and the BEC now would melt into hot boson gas. Thus the resultant structure would be the mixture of the central black hole and the surrounding hot boson gas.

In any way, the universe becomes very clumpy at small scales. The collapsed BEC will gravitationally attract baryons to form a cluster, as in the standard CDM model but in much smaller scales. On the other hand at large scales, the potential is not affected and remains the same as the standard $\Lambda$CDM model. This point will be further clarified later.

According to the above scenario, black holes may be common objects which are made from Dark Energy.

# 6. Observational tests for the model

The above scenario of BEC cosmology may appear to be quite different from the standard $\Lambda$CDM model. However this is not true. The BEC model can be designed to have maximum accordance with the standard model as we will see now.

*Large scale structure*

The collapse of the condensate proceeds in the smallest scale, as is seen from Eqs. (25)(33). This collapsing time scale is too short compared to the characteristic time scale of the larger scale fluctuations. In the linear stage of the density fluctuations, therefore, the influence of the BEC collapse does not show up in the larger scales. Once the uniform BEC is collapsed into compact objects such as boson stars and black holes, they work as the cold dark matter (CDM). Therefore, the global universe evolves as the same way as the standard model. Especially, BEC model does not affect the observed power spectrum $P(k)$ of density fluctuations in the linear (large scale) regime though it extremely enhances the smaller non-linear regime.

Let us further examine why the power spectrum is modified only in the smallest scale. As is shown in Eqs.(25)(33), the condensation collapses in a very short time although the evolution speed is huge. Therefore the large scale fluctuations simply do not have time to grow. Suppose the condensation collapses at the scale $l = a/k \equiv \tilde{k}^{-1} \ll H^{-1}$. Then, for the power law expansion $a \approx t^{2/3}$, the fluctuation in k-mode behaves as $\delta_k \propto a^{2\tilde{k}/H} = t^{4\tilde{k}/(3H)}$. The time duration of the collapse, i.e. the time necessary for the condensation to disappear, would be $\Delta t \approx l/c$ as estimated in Ref.[3]. Note that all the fluctuations are extraordinary unstable only within this time. During this time duration, the linear fluctuation of the scale $l$ can grow

$$\frac{\delta(t+\Delta t)}{\delta(t)} = (1+\frac{H}{\tilde{k}})^{4\tilde{k}/(3H)} \approx e^{4/3} \approx 3.8 \qquad (36)$$

at most. Moreover the larger scale fluctuation than $l$ grows much smaller. For example, the fluctuation at the scale $10 \times l$, the growing rate is only $\delta(t+\Delta t)/\delta(t) \approx 1.14$.

The above mechanism for the linear fluctuations can be summarized as, *exceedingly violent instability at small scales does not leave trace in the large scales within its too short time interval.*

*Microwave Background Radiation*

The condensation is supposed to collapse well after the decoupling time $t_{dec} \approx 3.8 \times 10^5 [\text{yr}]$, which requires
$$t_{dec} < \Gamma^{-1}. \tag{37}$$
This is because the violent collapse of BEC and formation of localized objects would strongly destroy the observed isotropy of cosmic microwave background, if occurred prior to the decoupling. If the condition Eq.(37) holds, then the temperature inhomogeneity generated within the decoupling era would not be altered from the calculation in the standard $\Lambda$ CDM model. However the integrated Sacks-Wolfe effect has a chance to modify the spectrum. The detailed analysis is urgently desired. Eq.(37) claims the bose gas is almost adiabatic and only small interaction can induce level transition to yield BEC.

*Many Black holes*

If the mass of the collapsing BEC exceeds the critical mass Eq.(34), $M > M_{critical}$, then the quantum pressure cannot prevent the collapse. Thus the black hole is inevitably formed. If the power spectrum of the produced black hole is appropriate, such black hole could be the seeds of proto–galaxies, and the black hole association to a galaxy would become inevitable. However unfortunately, we have no reliable calculation method to obtain such power spectrum of black holes at least at present. This difficulty is accelerated by the complication mentioned around Eq.(35); BEC can melt into non-degenerate bose gas when the black holes are formed. The detail of the final structure is not known since we cannot calculate the complicated dynamics with shock waves, at least at present.

On the other hand, the above rough scenario naturally suggests that the black holes should have been formed in the very early stage of the cosmic expansion, and if the black hole attracts baryons around them, then they can form quasars which could ionize the universe. Thus, there is a possibility that quasars shone before stars.

# 7. Summary of BEC Cosmological model

We have examined the cosmological model in which Bose-Einstein condensation (BEC) is identified as Dark Energy (DE). Boson gas, originally behaving as CDM, condensates to form DE. Increased BEC density eventually collapses to form black holes and localized objects very rapidly and BEC disappears. The new condensation initiates until it collapses again. The universe is filled with such many black holes and localized objects. The universe finally enters into the inflationary regime. Since DM is formed as a collapse of DE at each time, self-organized critical (SOC) state is approximately established to yield DM/DE $\approx 1$. These rapid collapses take place very fast and well after the photon-decoupling. Therefore, no strong violation is expected in the linear results (power spectrums in the large scale structure and the cosmic microwave background) of $\Lambda$ CDM model.

In this model, we proposed four main mechanisms which form the backbone of cosmology. It should be emphasized that all of these are based on a very simple feature of the model; the steady slow BE-condensation process.

(a) We first proposed a novel mechanism of inflation which is supported by the steady slow BE-condensation process. Contrary to the ordinary inflationary scenario, it is free from the slow-rolling condition, and naturally induce inflationary expansion at late stage of the cosmic evolution irrespective of the detail of the initial conditions and the potential form. In this sense, this inflation is robust and ubiquitous. For example, this inflation mechanism can be applied to the one in the very early stage with different kind of boson field condensation. In this case, the universe must exit the inflationary regime to evolve further into the present stage. In this context, we point out that the both hand sides of the balance $\dot{V} = \Gamma \rho_g$, to guarantee the inflation, become exponentially small, and this balance gradually becomes very subtle in the course of cosmic expansion. Therefore, any small perturbation would easily destroy the balance. Then the condensate simply roles down the potential and begins to oscillate. Thus the universe exits from the inflationary regime.

(b) Then we proposed a natural solution for the cosmic coincidence ('Why Now?') problem. Since DE is transformed into DM through subsequent BEC collapses, it would be natural to expect that the amount of DE is almost the same as that of DM. Thus our model realizes the self-organized critical state in the cosmic history. However, this argument is quantitatively crude, and we must note that some other special coincidence would be necessary to explain exactly the present ratio of DE and DM.

(c) We proposed the very early formation of highly non-linear objects through BEC collapses. For example in the numerical demonstration, the first black holes were formed at z=17 (**Figure 8**). If they successfully attract baryons to form accretion disks, then they may be quasars which shone for the first time in the universe.

(d) We finally proposed log-z periodicity in the subsequent BEC collapsing time. This periodicity stems from the two basic facts; the source of BE-condensation $\rho_g$ exponentially reduces and there exists a maximum allowed BEC strength $\rho_\phi^{cr} = V_{\max} = m^4/(-2\lambda)$. For example in the numerical demonstration, BEC collapse takes place at $z = 17.0, 10.9, 7.1, 4.1$, which is almost log-periodic (**Figure 8**).

We would like to comment on the relationship between our cosmic BEC model and the laboratory BEC experiments. In laboratory experiments, variety of BEC is observed in dilute gases of alkali atoms[10]. They show quantized vortexes in quantum turbulence, etc. Among such interesting features, we would like to focus on the 'boson-nova'. After a stable BEC is once formed, we rapidly switch its pressure from positive to negative. Then BEC collapses $10^{-5}$ times denser only after 5ms and shows burst explosion (boson-nova) [11][12]. This set of collapse and the bounce repeats many times and even jet like emission is observed. Does the same feature exist also in the cosmic BEC? What is then the jets ejected from black holes? In the future, we would like to explore the cosmology inspired by the condensed matter laboratory physics, exactly as we did in the last two decades in particle Physics.

We conclude our paper by briefly describing the important related works to our model.
1. We have constructed our model without specifying the boson species, but eventually we have to identify it in order to examine the detail of the BEC model. The boson must be lighter than 2 eV or should have ultra-low temperature independent from the cosmic baryon-photon temperature. The axion satisfies both the conditions though its weak attractive interaction, which is necessary in our BEC model, is not clear. X-rays from the Sun has been examined in ref[13] as a possible phenomena related with the boson field. In this paper, the authors examined the data of Yohkoh solar X–ray mission and tried to find signals from radiative decays of new massive neutral particles. More likely possibility would be the fermion-pair condensation in the same way as the superconductivity and the superfluidity. In this case, the original fermion and the condensate boson can have different basic properties and interactions. This kind of possibility is examined in ref.[14].
2. Early formation of black holes was also proposed in ref.[15]. Closed domain walls have a possibility to form primordial black holes after the inflation in the early universe. On the other hand in our BEC model, the BEC collapse and the subsequent black hole formation are supposed to take place after the photon decoupling. Therefore if we obtain sufficiently small condensation rate $\Gamma$, we can avoid severe constraint from the isotropy of the cosmic background radiation.
3. This condensation rate $\Gamma$ can be calculated from the microscopic physics in principle. Because the level transition toward the ground state is necessary for the BEC condensation, and because the cosmic process is almost adiabatic, this quantity can actually be very small. Since this smallness should be the essence in our model to guarantee the consistency with the CMB isotropy, we need to calculate the actual value in the near future.

4. The constancy of the condensation rate $\Gamma$, the approximation we used, would be justified as follows. The essential process in our model is the very slow condensation of the boson gas into BEC. Initially in the over-hill regime, the former dominates the latter ($\rho_\phi \ll \rho_g$) and the one-way process from gas to BEC dominates. Since there is no relevant reverse-process from BEC to gas, we can approximate the transition rate is simply constant. On the other hand in the deep in the mini-inflationary regime, the BEC temperature becomes quite low and its evaporation to gas would be negligible. We only have to take care at the transition point where the rate $\Gamma$ would not be constant. We would like to include this elaboration in our future publications. Similar justification is widely used in various practical applications such as in ref.[16].
5. The relevance of the relativistic GP equation was also pointed out in ref.[17] where spintessence model for DE and DM were studied. In this connection, we have to say that our relativistic Gross-Pitaevski equation (7) is still a proposal, and we cannot exclude the possibility of other form.
6. The unification of DE/DM is a natural direction of research. For example, tachyonic scalar field was used to describe both DE/DM in ref.[18]. Another unification model of DE/DM, generalized Chaplygin gas model, was discussed in ref.[19]. These models seem to unify DE and DM more directly than in our BEC model.
7. For the self-gravitating BEC, numerical solutions were examined in ref.[20], and DE collapse in ref.[21] using several forms of potential. Reconsiderations and the integration of them would improve our present model, on which we would like to report in the near future.
8. Interaction between DM and DE would also be a natural direction of research. This approach leads to a similar set of equations as ours Eq.(22) or Eq.(26) but with different coupling form. For example, coupled quintessence model is discussed in ref.[22], in which a full set of critical /fixed points are derived, and the coupling is strongly constrained by CMB. On the other hand in our BEC model, since baryons do not enter into the interaction, the interaction strength is not constrained from CMB in this context. Phenomenological approaches toward the coincidence problem are in ref.[23], in which the desired condition that DE/DM ratio becomes constant deduces a special coupling strength between DE/DM. The essential difference between these interacting models and our condensation model is that the fundamental process in the former is the reversible interaction while in the latter it is the irreversible condensation. This point is reflected, for example, in the form of the coupling; the transition rates are proportional to $\dot\phi$ in the former and $\Gamma$ in the latter. This tiny difference in appearance causes an essential difference in the inflationary behavior. By the way, the evaporation process of the condensate, which is simply a reverse process of our condensation process, is treated as latter in ref.[16] as it should be.
9. We have proposed a novel mechanism of inflation or of accelerated expansion. The essence of the mechanism is the balance of condensation flow and the potential force. Because this condensation flow is an irreversible dynamics and therefore cannot be expressed in any form of potential. On the other hand similar mechanism of inflation is recently proposed in ref.[24]. In this scenario of chameleon cosmology, the effective potential becomes a summation of the term which is proportional to the matter density and the term of the ordinary exponential potential. The minimum of the effective potential therefore becomes time dependent. Contrary to this model, we emphasize that our mechanism of inflation is totally free from the choice of the potential provided the mass of the boson is larger than the Hubble cosmic expansion rate.
10. What happens to the future of the universe in this model? Will it inflate forever? The key condition for the inflation to continue forever is the balance Eq.(28) $\dot V = \Gamma \rho_g$. Since both sides of this balance reduces exponentially in time, $\propto e^{-\text{const}\cdot t}$, a slight disturbance could destroy this delicate condition. Once this balance is destroyed, the BEC oscillates, damps, and the inflation ceases. On the other hand in the inflating universe, no inhomogeneity can grow, and therefore at least inhomogeneity cannot destroy this balance. If we could successfully describe the graceful exit from this balance, we can apply this novel inflation mechanism also in the first inflation in the early universe.

We would like to improve our BEC model in relation with the above researches.

# Acknowledgements

T.F. is grateful to M. Khlopov for discussions. M.M. would like to thank Masahiko Okumura for useful discussions on BEC and providing various references. The work of T.F. is supported in part by the Grant-in-Aid for Scientific Research from the Ministry of Education, Science and Culture of Japan (#16540269).